\documentclass[twocolumn,showpacs,preprintnumbers,amsmath,amssymb]{revtex4}

\usepackage{graphicx}
\usepackage{dcolumn}
\usepackage{bm}
\def\mib#1{\mbox{\boldmath $#1$}}

\begin{document}


\title{
Non-ideal behavior of intramolecular structure factor of
dilute polymers in a theta solvent}

\author{Kenji Shimomura}
\author{Hiizu Nakanishi}
\affiliation{
Department of Physics, Kyushu University 33, Fukuoka 812-8581, Japan}

\author{Namiko Mitarai}%
\affiliation{
Niels Bohr Institute, University of Copenhagen, Blegdamsvej 172100,
Copenhagen, Denmark }%

\date{\today}

\begin{abstract}
We study the configurational properties of single polymers in a theta
solvent by Monte Carlo simulation of the bond fluctuation model.
%
%
The intramolecular structure factor at the theta point is found to be
distinctively different from that of the ideal chain.  The structure
factor shows a hump around $q\sim 5/R_g$ and a dip around $q\sim 10/R_g$
in the Kratky plot with $R_g$ being the radius of gyration.  This
feature is apparently similar to that in a melt.
The theoretical expression by the simple perturbation expansion to the
first order in terms of the Mayer function can be fitted to the obtained
structure factor quite well, but the second virial coefficient cannot be
set to zero.
\end{abstract}


\maketitle

\section{Introduction}

One of the basic premises in the polymer physics is that the bonds
connecting neighboring monomer units are uncorrelated beyond the
persistent length along the chain \cite{DeGennes}.  This property allows
us to consider only a flexible chain as long as we are interested in
large scale properties of a polymer chain that is much longer than its
persistent length.  In this sense, it was a little embarrassing to
realize that there is actually a long range correlation in the bond
orientation of a polymer chain, and that the bond-bond correlation
function decays not exponentially but as the power law
\cite{Schaefer1999,Schaefer2004,Wittmer2004}.

The traditional picture for the bond-bond correlation is based on a simple
calculation for the polymer with a fixed bond angle around freely
rotating bonds.  In this case, one can calculate the bond-bond
correlation function explicitly to show the exponential decay with a
persistent length \cite{Flory}.  The existence of the long range
correlation, however, has been pointed out
\cite{Schaefer1999,Schaefer2004}, and it was demonstrated recently that
the power law behavior is induced in the bond-bond correlation through
the interaction between monomers separated by a long distance in the
curvilinear coordinate along the chain  \cite{Wittmer2004, Wittmer2007,
Wittmer2007EPL, Beckrich2007, shirvanyants2008}.  The power law in the
bond-bond correlation holds not only for an excluded volume chain, but
also for a chain in a melt and in a theta solvent, where a polymer chain
is supposed to behave as an ideal chain \cite{Flory,DeGennes}.  This has
been confirmed both by numerical simulations and by theoretical
analyses.

This deviation from the ideality of a chain in a melt has been seen also
in the intramolecular structure factor.  For the ideal chain, the
structure factor decays as $q^{-2}$ in the intermediate range, $1/R_g
\lesssim q \ll 1/a$, where $R_g$ and $a$ are the gyration radius and the
bond length, respectively.  This $q^{-2}$ decay comes from the fractal
dimension of the ideal chain configuration.  The intramolecular structure
factor of the polymer chain in a melt has been studied numerically and
theoretically, and it has been found that there exists a substantial
deviation from the ideal chain behavior \cite{Beckrich2007}.

In this paper, we study the structure factor of a single polymer
molecule in a theta solvent, that is another situation where a polymer
chain is supposed to become ideal effectively. In a melt, the
interaction between monomers is screened by the existence of other
polymer chains, and the excluded volume effect is canceled exactly by
the induced attraction due to the incompressibility of the system
\cite{Flory,DoiEdwards} while the interaction in a theta solvent is
being adjusted by some fine tuned external parameter, such as
temperature, so that the excluded volume effect is compensated by the
attractive part of the interaction.  We study how this fine tuning of
the parameter may affect on the virtual ideality of the structure
factor.

This paper is organized as follow.  After quickly reviewing how the long
range correlation comes into the bond-bond correlation in Sec.2, the
model and the method of our simulations are described in Sec.3, and the
simulations results are given in Sec.4.  The theoretical analysis is
outlined in Sec.5 and the results are discussed in connection with those
for melt in Sec.6.  Detailed expressions of the theoretical analysis are
given in Appendix.

\section{Bond-bond correlations in a polymer chain}

Let us quickly review how the long range correlation should arise in the
bond-bond correlation along a polymer chain \cite{Wittmer2004}.  We
consider a single polymer that consists of $N$ monomers.  Let $\mib r_n$
($n=1,\cdots, N$) be the position of the $n$-th monomer in it, and the
bond vector is denoted as
\begin{equation}
  \mib a_n \equiv \mib r_{n+1}-\mib r_n .
\label{bond-vector}
\end{equation}
We define a subchain as a part of the chain, and
introduce the subchain vector as
\begin{equation}
\mib R_n(s) \equiv \mib r_{n+s}-\mib r_n = \sum_{r=n}^{n+s-1}\mib a_r.
\end{equation}

Now, we assume that the bond-bond correlation depends only on the chemical
distance between the bonds:
\begin{equation}
P(s) = {1\over a^2} \left< \mib a_n \cdot \mib a_{n+s}\right>,
\label{bond-corr}
\end{equation}
where $a$ is the average bond length and $\left<\cdots\right>$ denotes
the ensemble average.  Then, the size of the subchain does not
depend on $n$, and we have
\begin{equation}
R(s)^2 \equiv \left< \mib R_n(s)^2 \right>
\approx a^2 \left(s  + \int_0^s dr (s-r) P(r) \right)
\label{R(s)-P(r)}
\end{equation}
in the large $s$ approximation.  This gives 
\begin{equation}
P(s) \approx {1\over a^2}\,{\partial^2\over\partial s^2} R(s)^2
\sim s^{-\omega}
\label{bond-corr-scaling}
\end{equation}
with $\omega=2-2\nu$ 
if the subchain size scales as
\begin{equation}
R(s) \sim s^\nu
\end{equation}
with the exponent $\nu\ne 1/2$.
We have $\omega\approx 0.824$
for the excluded volume chain,  where $\nu\approx 0.588$.

For the ideal chain with $\nu=1/2$, Eq.(\ref{bond-corr-scaling}) gives
$P(s)=0$ as it should, but
for the case of apparent ideality of a polymer chain in a melt or a theta
solvent,
we have
\begin{equation}
P(s) \sim s^{-3/2}
\end{equation}
because of  the correction term,
\begin{equation}
R(s)^2 \approx a^2 s \left( c_0+c_1\, s^{-1/2}+ \cdots \right).
\end{equation}

\section{Model and Simulation Method}
We perform Monte Carlo simulations of the bond fluctuation model (BFM)
on the three dimensional cubic lattice \cite{Carmesin1988,Deutsch1991}.
A polymer chain consists of $N$ monomers, and the $i$-th monomer is
located at the center of a cubic cell $\mib r_i$, occupying the cell
with 8 vertices at lattice sites.  The bond length between consecutive
monomers along the chain should be in the range $[2,\sqrt{10}]$ with the
exception of $\sqrt 8$.  The $i$-th and $j$-th monomers that are not
consecutive along the chain, namely $j\neq i\pm 1$, interact each other
through the ``quasi-Lennard-Jones'' potential energy \cite{Ivanov1998}
\begin{equation}
{U(r_{ij})\over k_B T} = \left\{\begin{array}{l}
-\beta \Bigl( 2(r_{ij}-2)^3 - 3(r_{ij}-2)^2+1 \Bigr)
\\
\hfill \mbox{for } r_{ij}=2, \sqrt 5, \sqrt 6, \mbox{ and } \sqrt 8 \\
0  \\ \hfill\qquad \mbox{otherwise }
\end{array}
\right. ,
\label{U(r)}
\end{equation}
where $r_{ij}=|\mib r_i-\mib r_j|$ is the distance between the
interacting monomers, and $k_B$ and $T$ are the Boltzmann constant and
the temperature, respectively.  The dimensionless parameter $\beta$ is
proportional to the inverse temperature and characterizes the
interaction \cite{com1}.

We perform Monte Carlo simulations, using Metropolis method along with
the slithering snake algorithm to accelerate the relaxation towards
equilibrium \cite{Wall1975}.  In our system, there is only one polymer
chain with $N$ monomer units.  One Monte Carlo step consists of $N$
trials of random displacement to one of the nearby sites for randomly
chosen monomers, followed by $N$ slithering snake trials.

The quantities we study are the radius of gyration $R_g$,
\begin{equation}
R_g^2 \equiv 
\left<{1\over 2N^2}\sum_{i=1}^N\sum_{j=1}^N (\mib r_i-\mib r_j)^2 \right>,
\label{R_g}
\end{equation}
the intrachain structure factor $S(\mib q)$,
\begin{equation}
S(\mib q) \equiv \left<
{1\over N}\sum_{i=1}^N\sum_{j=1}^N e^{i\mib q\cdot(\mib r_i-\mib r_j)}\right> ,
\label{S(k)}
\end{equation}
and the bond-bond correlation $P(s)$ averaged over the chain,
\begin{equation}
P(s)\equiv  {1\over a^2 (N-1-s)}
   \sum_{i=1}^{N-1-s}\left< \mib a_i\cdot \mib a_{i+s}\right>
\label{def-P(s)}
\end{equation}
with the average bond length $a$,
\begin{equation}
  a^2 \equiv \left< {1\over N-1}\sum_{i=1}^{N-1} \mib a_i^2 \right>,
\label{a}
\end{equation}
where the angular brackets represent the ensemble average.

{Note that the radius of gyration can be expressed using the
bond-bond correlation by a similar equation as Eq.(\ref{R(s)-P(r)}):
\begin{equation}
R_g^2 \approx {1\over 6}a^2 N  \left[
          1+\int_0^N dr \left(1-{r\over N}\right)^3 P(r)
                               \right] .
\label{R_g-P}
\end{equation}

\section{Results}

\subsection{Theta Point}
\begin{figure}
\centerline{
\includegraphics[width=8cm]{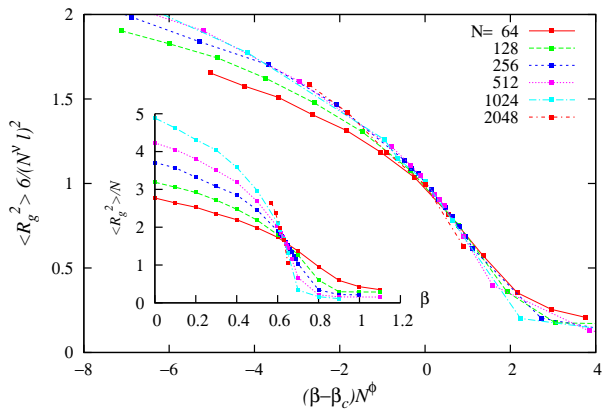}
}
\caption{(Color online)
The scaling plot for the radius of gyration $R_g$.
The scaled radius of gyration $R_g^2\, 6/(N^{\nu} \ell)^2$  are plotted
against $(\beta-\beta_c)N^\phi$ with $\beta_c=0.63$, $\nu=\phi=$ 0.5,
 and $\ell=3.17$ for various values of $N$
The inset shows the original plot of $R_g^2/N$ vs $\beta$.
}
\label{fig:R_g}
\end{figure}

First, we have to determined the theta point for our model.  It is
often defined as the point where the second virial coefficient vanishes.
In our simulations,  we define the theta point as the point where
the radius of gyration $R_g$ behaves as an ideal chain,
\begin{equation}
R_g \propto \sqrt N,
\label{R_g-ideal}
\end{equation}
in the large $N$ limit.  The interaction parameter $\beta_c$
at the theta point is determined numerically from the data of $R_g$ for
various values of $\beta$ and chain length $N$ by fitting them
to the finite size scaling form
\begin{equation}
R_g^2 = {1\over 6}N^{2\nu} \ell^2\, 
f \left( ({\beta-\beta_c)  N^\phi} \right) 
\label{R_g-N}
\end{equation}
with the two exponents, $\nu$ and $\phi$. The function $f(x)$ is the
scaling function that satisfies $f(0)=1$, and $\ell$ is the length scale
proportional to the average bond length.  Fig.\ref{fig:R_g} shows
that the data fit well to the scaling form (\ref{R_g-N})  with
\begin{equation}
\beta_c =0.63, \quad \nu=\phi=1/2, \quad \ell =3.17.
\end{equation}

\subsection{Theta Point determined by Bond-Bond Correlation}
\begin{figure}
\centerline{
\includegraphics[width=8cm]{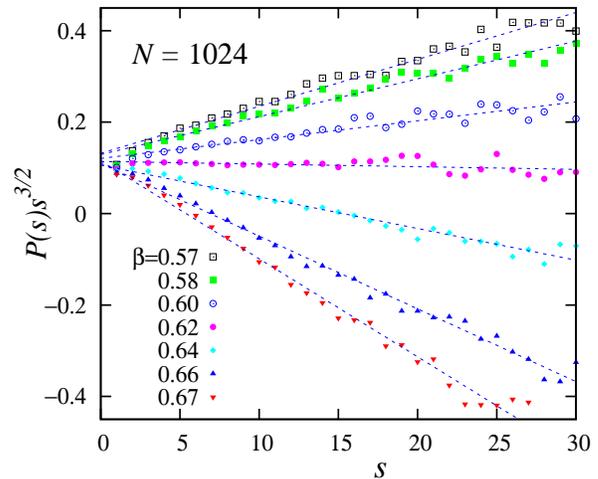}
}
\caption{(Color online)
Bond-bond correlation.
$P(s) s^{3/2}$ are plotted against $s$ for $N=1024$ for various $\beta$.
}
\label{fig:bond-corr}
\end{figure}
\begin{figure}
\centerline{ \includegraphics[width=8cm]{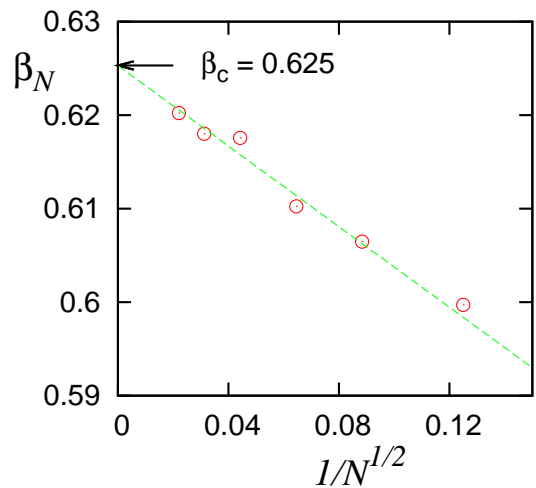} }
\caption{(Color online)
The theta point determined by the bond-bond correlation.  The values
of $\beta$ where the bond-bond correlation shows the ideal behavior,
$s^{-3/2}$, are plotted against $1/N^{1/2}$.  The linear extrapolation
to the infinite $N$ shows the theta point, $\beta_c=0.625$.}
\label{fig:beta_c}
\end{figure}

As we have discussed, the bond-bond correlation function decays as
$s^{-3/2}$ at the theta point.  Fig.\ref{fig:bond-corr} shows the
bond-bond correlation function $P(s)$ for various temperatures near the
theta point for $N=1024$.  It shows that the data fit to the form
\begin{equation}
P(s) s^{3/2}  = \tilde B_0\Bigl( 1-\beta/\beta_c(N)\Bigr) s + \tilde A + \cdots
\label{P(s)}
\end{equation}
that is consistent with the theoretical results by Shirvanyants, {\it et
al.}  \cite{shirvanyants2008}.  The theta point may be defined as
$\beta=\beta_c(N)$ where the bond-bond correlation decays as $s^{-3/2}$,
but it turns out that $\beta_c(N)$ depends on $N$.  The $N$-dependence
of $\beta_c(N)$ plotted in Fig.\ref{fig:beta_c} shows
\begin{equation}
\beta_c(N) - \beta_c \propto {1\over\sqrt N}
\label{beta_c(N)}
\end{equation}
with the theta point in the infinite $N$ limit $\beta_c=0.625$.  This value
is close enough to the previous estimate of $\beta_c=0.63$ by the
scaling plot of $R_g$.
In the rest of the paper, we use $\beta_c=0.63$ for the theta point.
Note that the $N$-dependence of $\beta_c(N)$ given by Eq.(\ref{beta_c(N)}) is
consistent with the ideal chain behavior of $R_g$ of Eq.(\ref{R_g-ideal}) at
$\beta=\beta_c$ in respect to Eq.(\ref{R_g-P}).

\subsection{Structure Factor}

\begin{figure}
\centerline{ \includegraphics[width=8cm]{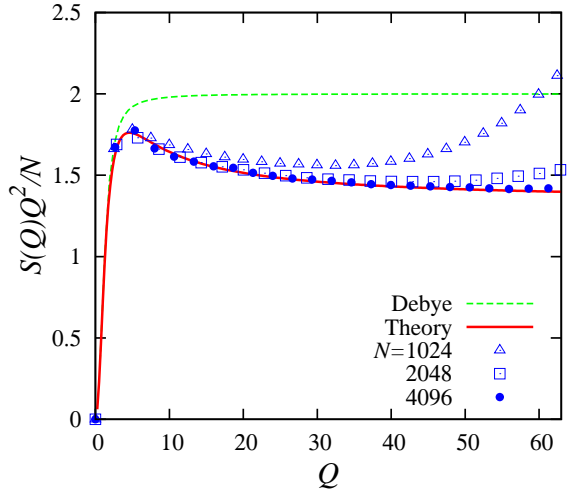} }
\caption{(Color online)
Kratky plots for the structure factor against $Q\equiv qR_g$.
The simulation data at $\beta=0.63$ for the chain with the length
$N=$1024, 2048, and 4096 are plotted along with the Debye scattering function
(the dashed line).  For the value of $R_g$, the values obtained by the
simulations are used for the simulation data and $R_{g0}=Na^2/6$ for
the analytical expressions.  The red curve represent the theoretical
estimate, Eq.(\ref{delta_S(q)}), with $\sqrt N B=-0.55$ and $A=0.154$.
} \label{fig:S_sim}
\end{figure}

The structure factor for the ideal chain, $S_0(q)$, in
the large $N$ limit is given by
\begin{equation}
S_0(q) = N f_D(qR_{g0})
\label{S_0}
\end{equation}
with the radius of gyration for the ideal chain
\begin{equation}
R_{g0}^2 \equiv {1\over 6} Na^2.
\end{equation}
and the Debye scattering function, 
\begin{equation}
f_D(x)\equiv {2\over x^4}\Bigl(e^{-x^2}-1+x^2 \Bigr).
\end{equation}

In the intermediate length scale, i.e., $1/R_{g0}\ll q\ll 1/a$, this decays as
\begin{equation}
S_0(q) \approx {12\over a^2}\,  q^{-2} + O(q^{-4}).
\end{equation}
The $1/q^2$-dependence comes from the scaling behavior of the ideal chain,
therefore, the existence of plateau in
the plot of $S(q)\, q^2$, i.e., Kratky plot,  has been considered to be a
sign of the ideality in a polymer chain behavior.

Fig.\ref{fig:S_sim} shows the Kratky plot of our numerical simulations
at the theta point for $N=1024\sim 4096$; The wave number $q$ is scaled
by the numerically obtained radius of gyration $R_g$ as $Q\equiv qR_g$.
As the number of monomers $N$ increases, the curve in the smaller $Q$
regime tends to converge to a common trend, but it is clearly different
from that of the ideal chain (the dashed line).  The Kratky plot at the
theta point shows a hump around $Q\sim 5$ and a dip around $Q\sim 10$,
and does not show the plateau as the one would expect for the ideal
chain.  Their general features are apparently similar to those found for
a polymer chain in a melt \cite{Beckrich2007}.

\section{Theory}

Now, we estimate the structure factor theoretically and compare it
with those obtained by the simulations.  Suppose $u(\mib r_{s,r})$ be the
interaction potential between the $s$-th and the $r$-th monomers with
$\mib r_{s,r}\equiv \mib r_s-\mib r_r$.  Then
the structure factor Eq.(\ref{S(k)}) is written as
\begin{equation}
S(\mib q) = {2\over N} 
\sum_{n>m} {\left<e^{i\mib q\cdot\mib r_{n,m}}e^{-U/k_B T}\right>_0
             \over \left< e^{-U}\right>_0 } + 1,
\end{equation}
where
\begin{equation}
U\equiv \sum_{r>s} u(\mib r_{r,s}),
\end{equation}
and $\left<\cdots\right>_0$ represents the statistical average for the
ideal chain.  

We now employ the approximation
\begin{equation}
e^{-U/k_B T} \approx 1 + \sum_{r>s} f(\mib r_{r,s})
\end{equation}
using the Mayer function
\begin{equation}
f(\mib r_{r,s}) \equiv e^{-u(\mib r_{r,s})/k_B T}-1.
\end{equation}
Then up to the first order in $f$, the correction in the structure factor
$\delta S(\mib q)$
is given by
\begin{eqnarray}
\delta S(\mib q) & \equiv & S(\mib q)-S_0(\mib q)
\nonumber \\
& \approx &
{2\over N}\sum_{n>m}\,\sum_{r>s}\Bigl[
\left< e^{i\mib q\cdot\mib r_{n,m}}f(\mib r_{r,s})\right>_0 
\nonumber\\ & & \hspace{5em}
-\left< e^{i\mib q\cdot\mib r_{n,m}}\right>_0\left<f(\mib r_{r,s})\right>_0 
\Bigr].
\label{delta_S-expand}
\end{eqnarray}

\begin{figure}
\centerline{
\includegraphics[width=8cm]{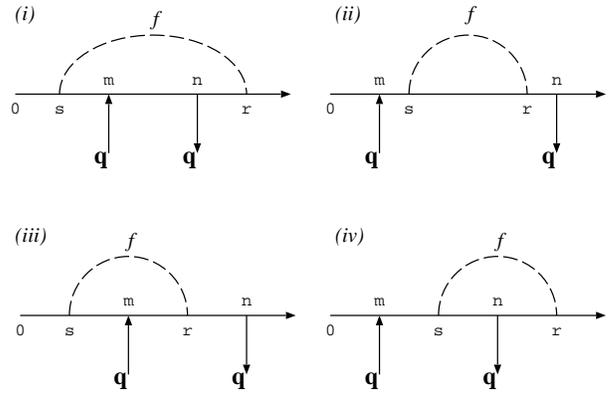}
}
\caption{Diagrams that represent contribution for Eq.(\ref{delta_S-expand}).
}
\label{Diagrams}
\end{figure}

There are four types of contribution in Eq.(\ref{delta_S-expand})
depending upon the relative positions of $n$, $m$, $r$, and $s$
(Fig.\ref{Diagrams}).  Adopting the bead-spring model for the ideal
chain average $\left<...\right>_0$ with the average bond length $a$, and
employing the further approximation valid for $qa\ll 1$, we obtain the
expression
\begin{equation}
{1\over N}\delta S(\mib q) \approx 
F(qR_{g0})\sqrt N B + G(qR_{g0}) A
\label{delta_S(q)}
\end{equation}
up to the second leading order in $N$.  The functional forms for $F$
and $G$ are given in Appendix.  The dimensionless parameters,
\begin{eqnarray}
B & \equiv & -{1\over a^3} \int d\mib r f(\mib r),
\label{B-def}
\\
A & \equiv & {1\over a^5} \int d\mib r\,  r^2 f(\mib r),
\label{A-def}
\end{eqnarray}
characterize the interaction.  The parameter $B$ is twice of the second
virial coefficient for the unlinked monomer gas \cite{Hansen}, and is
supposed to be close to zero for the theta solvent \cite{DoiEdwards}.
Note that $B$ comes into $S(q)$ as $\sqrt N B$.

\begin{figure}
\centerline{
\includegraphics[width=8cm]{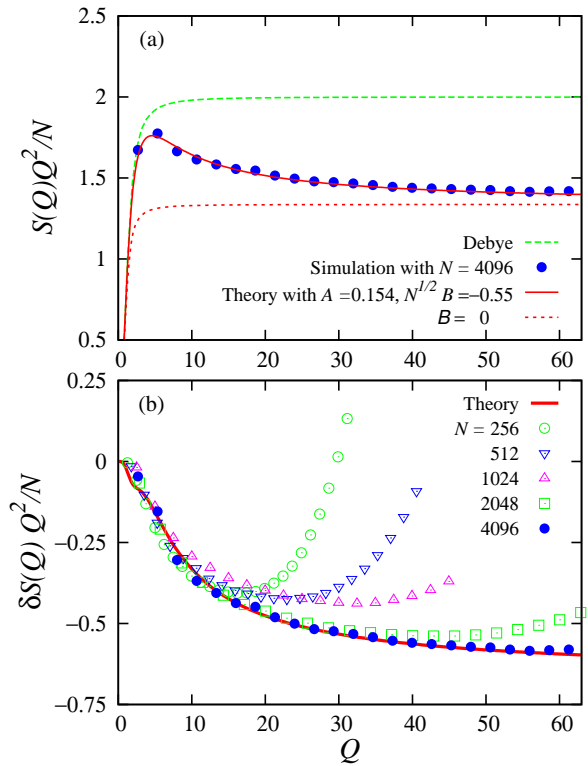}
}
\caption{(Color online)
Theoretical results for structure factors.
The numerical estimates by Eq.(\ref{delta_S(q)}) are plotted 
 with $A=0.154$ and $\sqrt N B=-0.55$, and 0.
}
\label{fig:S_theor}
\end{figure}
\begin{figure}
\centerline{
\includegraphics[width=8cm]{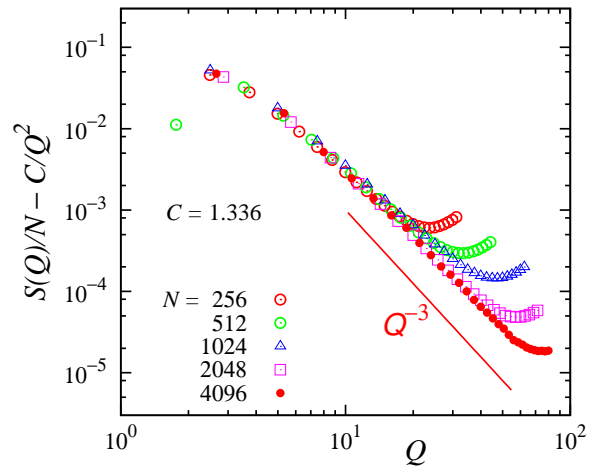}
}
\caption{(Color online)
The large $Q$ behavior of the structure factor after subtracting $C/Q^2$
with $C=1.336$.
}
\label{fig:S_asym}
\end{figure}

Fig.\ref{fig:S_theor}(a) shows the Kratky plots for the theoretical
structure factor $S(Q)Q^2$ with $Q\equiv q R_{g0}$ by
Eq.(\ref{delta_S(q)}) with $A=0.154$ and $\sqrt N B=-0.55$ along with
the curve for $A=0.154$ and $B=0$, the Debye function, and the
simulation data for $N=4096$.  One can see that the curve for $B=0$ is
almost proportional to the Debye function and cannot be fitted to the
simulation data.  Fig.\ref{fig:S_theor}(b) shows $\delta S(Q) Q^2/N$
with the simulation data for various values of $N$.  The data converge
to the theoretical curve quite well as $N$ increases.

In Fig.\ref{fig:S_asym}, the large $Q$ behavior of the structure factor
are plotted in the logarithmic scale, after subtracting the leading
order term of $C/Q^2$ with $C=1.336$.  It shows clearly that the second
leading order is $1/Q^3$, which is consistent with the asymptotic
expression we obtained in Eq.(\ref{A.delta_S-large-Q}).

\section{Discussions}

Our findings are summarized as follows:
(i) The theta point for the finite chain $\beta_c(N)$, where the
bond-bond correlation decays as $s^{-3/2}$, depends on the chain length
$N$, and in the infinite chain length limit it converges to the theta
point $\beta_c$ determined by the scaling behavior of the radius of
gyration.
(ii) The structure factors at the theta point $\beta_c$ obtained by the
numerical simulations are distinctively different from that of the ideal
chain, i.e.,
(ii-a) in the intermediate range of $q\gtrsim 1/R_g$, Kratky plot of the
structure factor shows a hump and a dip, and
(ii-b) in the larger $q$ range, $1/R_g\ll q \ll 1/a$, the structure
factor decays as $1/q^2$ with the next order term of $1/q^3$.
(iii) In the large $N$ limit, numerically obtained structure factors fit
well to the simple perturbation expression up to the first order of the Mayer
function with the fitting parameters $\sqrt N B=-0.55$ and $A=0.154$.

In comparison with the case of a melt, the obtained structure factors at
the theta point apparently resemble the ones of a melt in the existence
of a hump and a dip  \cite{Beckrich2007}.  In the large $q$ region,
however, the structure factors for the two cases differ; In the case of
melt, it has been shown by numerical simulations that structure factors
decays as $1/q^3$ for the large $q$ region  \cite{Beckrich2007}, in
contrast with the present case, where we found the $1/q^2$ decay with
the positive $1/q^3$ correction.  The $1/q^3$ decay in the melt was
interpreted as a result of renormalization from the $1/q$ term obtained
by the one-loop approximation \cite{Beckrich2007} while the
$1/q^2$ term and the $1/q^3$ term at the theta point directly correspond
to the $A$-term and the $\sqrt NB$-term, respectively, in our
theoretical expression.
In the real space, the $1/q^2$ decay of the structure factor in large $q$
means the $1/r$ density correlation of the ideal chain in short $r$.
The existence of the positive $1/q^3$ correction and the hump around
$q\sim 5/R_g$ implies that the density correlation does not decay as fast
as that of the ideal chain around $r\sim R_g$.

The simulated structure factors here at the theta point resemble those
of a melt, but it is intriguing that we need to set $\sqrt N B\ne 0$ for
our theoretical expression of the structure factor to fit to the
numerical results at the theta point $\beta_c$.  The theta point is
often considered as the point where the second virial coefficient $B/2$
vanishes.  Actually, if we set $B=0$, the structure factor shows Kratky
plateau and looks pretty much like the ideal one (the dotted line in
Fig.\ref{fig:S_theor}(a)), but this cannot be fitted to the simulation
results by adjusting $A$ alone even with an arbitrary factor.
Theoretically,
if we consider higher order corrections, the theta point should 
correspond to the vanishing point of the interaction parameter $z$
\begin{equation}
 z \equiv \sqrt N b
\end{equation}
where $b$ is the excluded volume parameter with the additive
renormalization by the higher order cluster contributions, and may be
given in the form
\begin{equation}
b = B + C N^{-1/2} + \dots
\label{ren-B}
\end{equation}
with a constant $C$.  However, such renormalization effect could be
partially taken into account by replacing $B$ with $b$, and one
would expect the structure factor at the theta point should be given by
the one with $\sqrt N B=0$ in our expression.  }

In the bond-bond correlation $P(s)$, we observed the analogous
$N$-dependence, which can be interpreted in the same way.  Shirvanyants
et al. \cite{shirvanyants2008} have obtained the theoretical expression
corresponding to Eq.(\ref{P(s)}) in a similar approximation to the one
we employed for $S(\mib q)$, and showed that the coefficients for the
$s^{-1/2}$ and the $s^{-3/2}$ terms in Eq.(\ref{P(s)}), i.e.  $\tilde
B_0(1-\beta/\beta_c(N))$ and $\tilde A$, are proportional to the
parameters $B$ and $A$, respectively.
The observed $N$-dependence in $\beta_c(N)$ as Eq.(\ref{beta_c(N)})
in our simulations should be a result of higher order effects, that
could be obtained by replacing the coefficient of $s^{-1/2}$-term with
the renormalized one as Eq.(\ref{ren-B}).

Regarding the origin for the deviation of the structure factor from the
Debye function, an obvious possibility could be that inaccurate estimate
for the theta point $\beta_c$, i.e. our estimate $\beta_c=0.63$ is not
close enough to the theta point for the structure factor to be
Debye-like even though the two independent estimates from the radius of
gyration $R_g$ and the bond correlation coincide with each other within
our numerical precision.  This might happen due to the slow convergence
caused by the logarithmic correction at the theta point.
Another possibility, perhaps more interesting one, would be that the
tricritical fluctuations at the theta point produce a non-trivial
contribution to the structure factor.
%


\begin{acknowledgments}
The authors thank Dr. Takahiro Sakaue for critical reading of the manuscript.
\end{acknowledgments}


\begin{widetext}
\appendix
\section{First oder calculation of $S(\mib q)$ in the Mayer function}

In the appendix, we describe the calculation of Eq.(\ref{delta_S-expand})
to obtain the explicit expressions for $F$ and $G$ in Eq.(\ref{delta_S(q)}).

The contribution from the diagram (i) in Fig.\ref{Diagrams} is given by
\begin{eqnarray}
S_i(\mib q)  & =  &
{2\over N}\sum_{1\le s<m<n<r\le N}
\Bigl[
  \left< e^{i\mib q\cdot\mib r_{n,m}} f(\mib r_{r,s})\right>_0
- \left< e^{i\mib q\cdot\mib r_{n,m}}\right>_0 \left< f(\mib r_{r,s})\right>_0 
\Bigr]
\nonumber \\ & \equiv &
{2\over N}\sum_{1\le s<m<n<r\le N} H(q; r-s, n-m),
\end{eqnarray}
where
\begin{equation}
H(q;l_1,l_2)  \equiv 
\int d\mib r_1 f(\mib r_1)
\int d\mib r_2 e^{i\mib q\cdot\mib r_2}
\Bigl(
G_0(\mib r_1-\mib r_2;l_1-l_2)-G_0(\mib r_1;l_1)
\Bigr)
G_0(\mib r_2;l_2)
\end{equation}
with the free propagator
\begin{equation}
G_0(\mib r; n) \equiv \left({3\over 2\pi a^2}\right)^{3/2}
\exp\left[ -\,{3\; \mib r^2\over 2a^2\, n}\right] .
\end{equation}
$H(q;l_1,l_2)$ can be estimated as
\begin{eqnarray}
\lefteqn{
H(q;l_1,l_2) =
e^{-(1/6)q^2 a^2 l_2}
   \left({3\over 2\pi a^2 l_1}\right)^{3/2}
}
\nonumber\\ & & \hspace{5em}
 \int d\mib r\; e^{-3 r^2/2a^2 l_1}
  \left(
    e^{(1/6)q^2 a^2 l_2^2/l_1}
    \cos\left[{{l_2\over l_1} \mib q\cdot \mib r } \right]
-1 \right)  f(\mib r)
\nonumber\\
&\approx &
 \left({3\over 2\pi}\right)^{3/2}
   \exp\left[-{1\over 6}q^2 a^2 l_2\right] {1\over l_1^{3/2}}
\Biggl[
-\left( \exp\left[{q^2a^2\over 6}{l_2^2\over l_1}\right]-1  \right) B
\nonumber\\ & & \hspace{1cm}
-\Biggl\{
 {3/2\over l_1}
 \left( \exp\left[{q^2a^2\over 6}{l_2^2\over l_1}\right]-1  \right)
 + {1\over 6}q^2 a^2 \exp\left[{q^2a^2\over 6}{l_2^2\over l_1}\right]
               \left({l_2\over l_1}\right)^2
 \Biggr\} A
\Biggr].
\end{eqnarray}
In the last expression, we have expanded the cosine up to $q^2$, which
is valid for $qa\ll 1$.

The contributions from (ii), (iii), and (iv) can be given in terms of
$H$ as
\begin{eqnarray}
S_{ii}(\mib q) & = &
{2\over N}\sum_{1\le m <s<r<n <\le N}\hspace{-3ex}
 e^{-(1/6)q^2 a^2 (n-r)}\; H(q;r-s,r-s)\; e^{-(1/6)q^2 a^2 (s-m)}
\\
S_{iii}(\mib q) & = &
{2\over N}\sum_{1\le s <m<r<n <\le N}
e^{-{(1/6)q^2 a^2 (n-r)}} H(q; r-s, r-m)
\\
& = & S_{iv}(\mib q).
\end{eqnarray}

For large $N$, we can replace the summation by integral and obtain,
to the leading orders in $N$,
\begin{eqnarray}
S_i(\mib q) & \equiv & N^{3/2} F_i(qR_{g0} ) B + N^{1/2} G_i(qR_{g0}) A,
\\
S_{ii}(\mib q) & \equiv & N^{3/2} F_{ii}(qR_{g0} ) B + N G_{ii}(qR_{g0}) A,
\\
S_{iii}(\mib q) & \equiv & N^{3/2} F_{iii}(qR_{g0} ) B + N^{1/2} G_{iii}(qR_{g0}) A,
\end{eqnarray}
with
\begin{eqnarray}
F_i(Q) & = &  -2\left({3\over 2\pi}\right)^{3/2}
\int_0^1 dp \int_0^1 dl \,\sqrt p (1-p) (1-l) 
       \left( e^{-Q^2 pl(1-l)} - e^{-Q^2 pl}\right),
\\
F_{ii}(Q) & = & -2\left({3\over 2\pi}\right)^{3/2}
\int_0^1dp \int_p^1 dl\,
{e^{Q^2p}-1\over p^{3/2}}\, (1-l)(l-p)e^{-Q^2l},
\\
F_{iii}(Q) & = & -2\left({3\over 2\pi}\right)^{3/2} 
\int_0^1 dp\int_0^1 dl
     \left( 1-p + {e^{-Q^2(1-p)}-1\over Q^2} \right)
\nonumber\\& & \hspace{13em}\times
    \left( {e^{-Q^2 pl(1-l)}-e^{-Q^2 pl}\over Q^2 \sqrt p} \right),
\end{eqnarray}\begin{eqnarray}
G_i(Q) & = &
-2\left({3\over 2\pi}\right)^{3/2} 
\int_0^1 dp \int_0^1 dl \sqrt p (1-p)(1-l) 
\nonumber \\ & & \hspace{8em}\times
\left[
{3\over 2} {e^{-Q^2 pl(1-l)}-e^{-Q^2 pl}\over p} +
Q^2 l^2 e^{-Q^2 pl(1-l)}
\right],
\\
G_{ii}(Q) & = &
-5\left({3\over 2\pi}\right)^{3/2} \zeta({3/2})
\left[   \left({2\over Q^4}+{1\over Q^2}\right) e^{-Q^2}
               -\left({2\over Q^4}-{1\over Q^2}\right)  \right],
\\
G_{iii}(Q) & = &
-2\left({3\over 2\pi}\right)^{3/2} {1\over Q^2}
\int_0^1 dp \int_0^1 dl {1\over \sqrt p}
\left[ 1-p + {e^{-Q^2(1-p)}-1\over Q^2} \right],
\end{eqnarray}
where
$\zeta(s)$ is Riemann's zeta function and $\zeta(3/2)=2.61237\cdots$.

With these functions, the correction of the structure function is given by
\begin{equation}
{1\over N} \delta S(\mib q) \approx
F(qR_{g0})\, \sqrt N B + G(qR_{g0})A
\end{equation}
with
\begin{eqnarray}
F(Q) & \equiv & F_i(Q)+ F_{ii}(Q)+2F_{iii}(Q),
\\
G(Q) & \equiv &  G_{ii}(Q)
\end{eqnarray}
in the leading orders in $N$.
For large $Q$, i.e., $1/R_{g0}\ll q \ll 1/a$, we have
\begin{equation}
{1\over N} \delta S(\mib q) \approx
-8\left({3\over 2\pi}\right)^{3/2}\left[
\left(\sqrt\pi + C \right) 
         {1\over Q^3}(\sqrt N B)
         +{5\over 8}\zeta(3/2){1\over Q^2} A
\right]
\label{A.delta_S-large-Q}
\end{equation}
with $C\approx 1.01171$, and for small $Q$, i.e., $q\ll 1/R_{g0}$, 
\begin{equation}
{1\over N} \delta S(\mib q) \approx
-2\left({3\over 2\pi}\right)^{3/2} 
\left( {59\over 315} \sqrt N B + {5\over 12}\zeta(3/2) A \right) Q^2
\label{A.delta_S-small-Q}
\end{equation}
with $Q\equiv qR_{g0}$.

From Eq.(\ref{A.delta_S-small-Q}), the correction for the radius of gyration
$R_g$ is obtained as
\begin{equation}
R_g^2 \approx
\left[
1+2
\left({3\over 2\pi}\right)^{3/2} 
\left( {59\over 105} \sqrt N B + {5\over 4}\zeta(3/2) A \right)
\right] R_{g0}^2 .
\end{equation}

\end{widetext}


\end {document}